\documentclass[11pt]{article}
\usepackage[a4paper,margin=2.5cm]{geometry}
\usepackage[T1]{fontenc}
\usepackage{tikz}
\usepackage[
        backend=biber,
	giveninits=true,
	style=numeric,
        sorting=nyvt,
        maxnames=5,
        maxalphanames=5,
        url=true,
	isbn=false
           ]{biblatex}
\renewbibmacro*{doi+eprint+url}{%
    \iftoggle{bbx:doi}{\printfield{doi}}{}%
    \newunit\newblock
    \iftoggle{bbx:eprint}{\usebibmacro{eprint}}{}%
    \newunit\newblock
    \iftoggle{bbx:url}{\iffieldundef{doi}{\usebibmacro{url}}{\clearfield{url}}}{}%
} 

\DeclareSourcemap{
  \maps[datatype=bibtex, overwrite]{
    \map{
      \step[fieldset=annotation, null]
      \step[fieldset=keywords, null]
    }
  }
}

\usepackage{tikz}
\usepackage{amsmath,amsfonts}
\usepackage{amssymb}
\def\diffd{\mathrm{d}}
\usepackage{hyperref}
\hypersetup{hidelinks}

\def\sign{\mathop{\text{sign}}\nolimits}
\def\erfc{\mathop{\text{erfc}}\nolimits}

\def\E{\mathbb E}
\usepackage{bbm}
\def\indic#1{\mathbbm1_{\{#1\}}}

\begin{document}

\title{The flux of particles in a one-dimensional Fleming-Viot process}
\author{\'Eric Brunet$^1$ and Bernard Derrida$^2$\\\footnotesize
$^{1,2}$~Laboratoire de Physique de l'\'Ecole Normale Sup\'{e}rieure, ENS Universit\'{e} PSL,\\[-.5ex]\footnotesize CNRS, Sorbonne Universit\'{e},    Universit\'{e} de Paris, Paris, France.\\[-.5ex]
\footnotesize $^2$~Coll\`{e}ge de France, Paris, France.}

\maketitle

\begin{abstract}{}
The Fleming-Viot process describes a system of $N$ particles diffusing on a graph with an absorbing site. Whenever one of the particles is absorbed, it is replaced by a new particle at the position of one of the $N-1$ remaining particles. Here we consider the case where  the particles lie on the semi-infinite line with a biased diffusion towards the origin which is the absorbing site. In the large $N$ limit, the evolution of the density becomes deterministic and has a  number of characteristics similar to the Fisher-KPP equation: a  one-parameter family of steady state solutions, dependence of the long time asymptotics  on the initial conditions, Bramson logarithmic shift, etc. One noticeable difference, however, is that in the Fleming-Viot case, the solution can be computed explicitly for arbitrary initial conditions and at an arbitrary time. By modifying the diffusion rule near the origin,  one can produce a transition in the flux of absorbed particles, very similar to the pushed-pulled transition in travelling waves. Lastly, when a cut-off of order~$1/N$ is added to the hydrodynamic equation, we find corrections of order $(\ln N)^{-2}$ or $N^{-\gamma}$ to the flux of absorbed particles, depending on whether the front is pulled or pushed, as for the correction to the velocity in travelling waves.

\end{abstract}

\section{Introduction}

The Fleming-Viot process can be viewed as a set of $N$ non-interacting particles performing random jumps on a graph with an absorbing vertex. Whenever a particle hits the absorbing vertex, it is instantaneously relocated at the position of one of the $N-1$ remaining particles chosen uniformly at random.
(In the original version \cite{FlemingViot.1979,EthierKurtz.1993} introduced in the context of genetics, all the sites were connected with a small rate to the absorbing vertex.)

 One question, then, is to know  whether the system reaches a steady state in the long time limit, and what are its characteristics~\cite{Villemonais.2015}.

On the same graph, one can also consider a single particle performing random jumps.  It is well established \cite{FerrariMaric.2007,Villemonais.2014,Villemonais.2015} that the distribution of the position of a single particle at time $t$ conditioned on not being absorbed \cite{FerrariMartinezSanMartin.1997} is the same as the large~$N$ limit of the density of particles in a Fleming-Viot process where the $N$ particles are initially independently distributed as the single particle. This hydrodynamic limit holds for any finite time~$t$.

A stationary distribution for the diffusion of a single particle conditioned on being still alive is called a quasi-stationary distribution (qsd).
For a finite connected graph, there exists a unique quasi-stationary distribution, and the density of the Fleming-Viot process for $N$ particles converges as $N\to\infty$ and $t\to\infty$ to that quasi-stationary distribution~\cite{AFG.2011,BHIM.1996,BHM.2000}. For an infinite graph, there might be infinitely many quasi-stationary distributions of the single particle problem, and a question is to determine which (if any)  of these quasi-stationary solutions is selected by a walker conditioned on surviving, depending on its initial distribution~\cite{Cavender.1978,FerrariMaric.2007,AFGJ.2016,ChampagnatVillemonais.2021,MartinezSanMartin.1994,MartinezPiccoSanMartin.1998}. Another question is to determine which quasi-stationary distribution is selected by the Fleming-Viot process as $t\to\infty$ and then $N\to\infty$. It is expected that, in a number of cases, the  selected solution by the Fleming-Viot process is the ``minimal'' one, which is defined as the quasi-stationary distribution for which the average time for a particle    to reach the absorbing state is minimum~\cite{Villemonais.2015,AFGJ.2016,GroismanJonckheere.2019,Tough.2025}.

This paper is organized as follows. 
In \autoref{sec:lattice}, we study the Fleming-Viot process consisting of $N$ particles performing continuous time random walks  biased towards the origin on  positive integers, with  $0$ as the absorbing site. According to the usual definition of Fleming-Viot processes, whenever a particle is absorbed, it is moved to the position of one of the $N-1$ remaining particles.
Here, we are interested in the total number $Q_t$ of particles absorbed up to time $t$, and in how this quantity  depends on the initial condition. In the $N\to\infty$ limit, $Q_t/N$ is directly related to the survival probability of a single walker. We obtain an explicit expression for this quantity, valid at any time~$t$ and for any initial condition.
This allows us to obtain its long time asymptotics;  we will show that it exhibits a behaviour similar to the position of a Fisher-KPP front in the large time limit, \cite{Fisher.1937,KPP.1937} including Bramson's logarithmic term~\cite{Bramson.1978,Bramson.1983}.

In \autoref{sec:transition}, we show that by modifying the jumping rates near the origin, the Fleming-Viot undergoes a transition very similar to  the pushed-pulled transition of travelling waves, including the same cross-over regime.~\cite{Derrida.2023}

In sections~\ref{sec:RL0} and~\ref{sec:RL1}, we  consider the Fleming-Viot in the continuum, which can be understood as a limiting case of the  discrete situation discussed in sections~\ref{sec:lattice} and~\ref{sec:transition}. In this continuum version, expressions are simpler, and one can analyze in more details the effect of initial conditions on the long time asymptotics and the transition from pulled to pushed fronts.

Lastly, in \autoref{sec:cut-off}, we determine for the Fleming-Viot problem the correction due to a cut-off~$1/N$ in its hydrodynamic equation. In the case of pulled front equations, the cut-off correction~\cite{BrunetDerrida.1997} is known to give the leading large $N$ correction to the velocity of the front for a $N$ particle stochastic system converging to that front equation as $N\to\infty$.
We find the same type of cut-off corrections for the Fleming-Viot problem and for front equations; in particular, in the pulled case, the flux of absorbed particles in presence of a cut-off of order~$1/N$ is increased by a quantity of order $(\ln N)^{-2}$ in the stationary regime as compared to its $N\to\infty$ limit.

\section{The Fleming-Viot process on positive integers}
\label{sec:lattice}
In this section we consider   a version of the Fleming-Viot process where the  $N$ particles  lie  on a one dimensional lattice with strictly positive integer positions. Each particle performs  a biased random walk in continuous time, independently of what the other particles do. If a particle is on site $j$ at time $t$, it moves during the  infinitesimal time interval $\diffd t$ to site $j+1$ with probability $\alpha \, \diffd t$ and to the site $j-1$ with probability $\beta \, \diffd t$. (The particle remains on site $j$ with probability $1-(\alpha+\beta)\diffd t$.) 
Whenever a particle hits the origin,  it is instantaneously relocated  at the position of one of the remaining $N-1$ particles chosen at random.
When $\alpha < \beta$, all the particles move towards the origin, and  the system reaches a steady state. (For  $\alpha \ge \beta$  the dynamics does not lead to a steady state; we do not discuss that case here and always assume $\alpha < \beta$).

From the definition of this process  it  is easy to see that the expectation $\langle n_j \rangle $ of  the number of particles $n_j$ on site $j$ at time $t$ evolves according to 
\begin{equation}
\frac{\diffd  \langle n_j \rangle}{ \diffd  t} = 
\alpha \langle n_{j-1} \rangle +\beta \langle n_{j+1} \rangle 
-(\alpha  +\beta )\langle n_{j} \rangle  + \frac\beta{ N-1} \Big( \langle n_1 n_j \rangle  - \langle n_1 \rangle \delta_{j,1} \Big)  . 
\label{eq:FV}
\end{equation}
with the convention that $n_0=0$. 
All the difficulty of the problem comes from the presence of the correlation $\langle n_1n_j\rangle$.

Let us now consider a single walker on the same lattice with the same transition rates, and let $v_j(t)$ be the probability that the walker has never reached the origin up to time $t$ and that it is at position $j$ at time $t$. Clearly, for $j\ge1$
\begin{equation}
\frac{\diffd  v_j}{\diffd t} = 
\alpha v_{j-1}  +\beta   v_{j+1} 
-(\alpha  +\beta )  v_{j}   \qquad \text{with }v_0=0.
\label{A10}
\end{equation}
Let $e^{-q_t}$ be the probability that the walker  has never reached the origin up to time~$t$:
\begin{equation}
e^{-q_t}=\sum_{k} v_k(t).
\label{C3}
\end{equation}
From \eqref{A10}, one gets
\begin{equation}
\frac{\diffd  e^{-q_t}}{\diffd t} = -\beta v_1.
\qquad\iff\qquad
\frac{\diffd q_t}{\diffd t} = \beta v_1 e^{q_t}=\beta u_1,
\label{B3}
\end{equation}
where $u_j$ is defined by
\begin{equation}
u_j(t)=  e^{q_t} v_j(t).
\label{A9}
\end{equation}
This quantity $u_j$ is the probability that the walker is at position $j$ at time $t$ \emph{given} that it has never reached the origin up to time~$t$. From \eqref{A10}, one obtains
\begin{equation}
\frac{\diffd   u_j}{\diffd t} = 
\alpha  u_{j-1}  +\beta   u_{j+1} 
-(\alpha  +\beta )  u_{j}  + \beta   u_1  u_j,
\label{A6}
\end{equation}
to be compared to \eqref{eq:FV}. 
In fact, if we make a mean-field approximation of \eqref{eq:FV}, which consists of replacing $\langle n_1      n_j\rangle$ by $\langle n_1\rangle \langle n_j\rangle$, and if we take $N\to\infty$, we obtain that $\langle n_j\rangle/N$ follows the same equation as $u_j$ in \eqref{A6}.

It is known \cite{FerrariMaric.2007,Villemonais.2014,Villemonais.2015} that for a fixed time $t$, if the initial positions of the $N$ particles in the Fleming-Viot are distributed independently according to $u_j(0)$, then
for all $N$ and $j$, one has
\begin{equation}
\frac{n_j(t)}N\to u_j(t)\quad\text{as $N\to\infty$}.
\label{eq:cvg}
\end{equation}
In other words, the mean-field theory of \eqref{eq:FV} becomes exact as $N\to\infty$.

Let $Q_t$ be the number of particles absorbed at the origin up to  time $t$. During $\diffd t$, it increases by one with probability $\beta n_1(t)\,\diffd t$, and so
\begin{equation}
\frac { \diffd  \langle Q_t \rangle }{ \diffd t} = \beta \langle n_1 \rangle  . 
\label{A4} 
\end{equation}
Dividing by $N$ and taking $N$ large, the right hand side converges to $\beta u_1=\beta v_1 e^{q_t}=\diffd q_t/\diffd t$, see \eqref{B3}, from which we obtain that
\begin{equation}
\frac{\langle Q_t\rangle}N
\to
q_t\qquad\text{as $N\to\infty$}.
\label{A7}
\end{equation}

\subsection{Steady state solutions}\label{sec:steadystate}

Let us first show that there exists a one parameter family of steady state solutions of \eqref{A6}. For a steady state solution, \eqref{B3} implies that
\begin{equation}
q_t= \kappa  t \qquad\text{with }\kappa=\beta u_1.
\end{equation}
The quantity $\kappa$ represents $\frac1N$ times the flux of particles being absorbed in the large~$N$ limit. Then, \eqref{A6} gives 
\begin{equation}
0=\alpha u_{j-1}+\beta u_{j+1}-(\alpha+\beta- \kappa)u_j\text{\quad for $j\ge1$},\qquad u_0=0.
\label{AA15}
\end{equation}
The solution to \eqref{AA15} with $u_0=0$ and $\kappa=\beta u_1$ is \cite{Maric.2015,vanDoorn.1991}
\begin{align}
u_j &= \frac{\kappa 
\left[ \left(\alpha + \beta - \kappa + \Delta \right)^{j} -\left(\alpha + \beta - \kappa - \Delta \right)^{j} \right] 
}{ (2 \beta)^j \,  \Delta}&&\text{if $0<\kappa<\kappa_c$},\label{A16}\\
u_j&=  \frac {(\sqrt{\beta} - \sqrt{\alpha})^2}{ \sqrt{\alpha\, \beta} }  \left(\sqrt{\frac\alpha \beta} \right)^{j} j && \text{if $\kappa=\kappa_c$}.
\label{A20}
\end{align}
where $\Delta$ and $ \kappa_c$ are given by
\begin{equation}
\Delta = 
\sqrt{(\alpha + \beta-\kappa)^2-4 \alpha \beta}, \qquad \kappa_c=(\sqrt{\beta} - \sqrt{\alpha})^2.
\label{eq:kappac}
\end{equation}
($\kappa_c$  is the smallest value of $\kappa$ for which $\Delta=0$.
There are no positive solutions if $\kappa\le0$ or $\kappa>\kappa_c$. One can check that $\sum_j u_j=1$.)

This is very reminiscent \cite{GroismanJonckheere.2019} of the Fisher-KPP equation \cite{KPP.1937,Fisher.1937}: there is a one parameter family of steady  state solutions, indexed by their flux $\kappa$, as there is a  one parameter family of travelling wave solutions in        Fisher-KPP, indexed by their velocity $v$. Under the condition that the solutions are positive, there is here a \emph{maximal} flux $\kappa_c$, and a \emph{minimal} velocity $v_c$ in Fisher-KPP. In both cases, asymptotically, the steady states or travelling waves decrease exponentially in space as in \eqref{A16} for $\kappa<\kappa_c$ or $v>v_c$, except that there is a linear prefactor for the extremal case as in \eqref{A20} for $\kappa=\kappa_c$ or $v=v_c$.

The question, then, is to determine which steady state  solution is reached for a given initial condition.

\subsection{The exact solution for arbitrary initial conditions}
\label{sec:solu}

The mean field equation \eqref{A6} for $u_j(t)$ can be solved exactly for an arbitrary initial condition $u_j(0)$. First, we need to solve
the linear equation \eqref{A10} with $v_j(0)=u_j(0)$ as initial condition, and then $u_j(t)$ is obtained from \eqref{C3} and \eqref{A9}.

To solve the linear equation \eqref{A10}, we first write
\begin{equation}
v_j(t) = \sum_{k \ge 1} \pi_{j,k}(t)   u_k(0),
\label{A12}
\end{equation}
where 
$ \pi_{j,k}(t)  $  satisfies the same evolution equation (\ref{A10}) as  $v_j(t)$, with the initial condition replaced by
$
\pi_{j,k}(0)  = \delta_{j,k} 
$.  
The quantity 
 $\pi_{j,k}(t)    $
is the probability  that  a random walker (jumping to the right at rate    $\alpha$ and to the left at rate $\beta$) goes from $k$ to $j$ in a time $t$ without visiting the origin.

One can check easily that
\begin{align}
\pi_{j,k}(t)&=\frac1{2\pi i}\oint \frac{\diffd z}z \biggl[z^j - \Bigl(\frac{\beta z}\alpha\Bigr)^{- j}\biggr]z^{-k}e^{\left(\beta z -\alpha-\beta+\frac\alpha z\right)t}
\label{A13}
\\&
= \frac1 \pi \int_{-\pi}^\pi \diffd \theta\,   
\left(\frac\alpha\beta\right)^{\frac{j-k}2}\sin(\theta j)\sin(\theta k)
\exp\left[ \left(2\sqrt{\alpha\beta} \cos\theta -\alpha-\beta\right) t \right],
\label{AA24}
\end{align}
where the contour in \eqref{A13} is any loop around $z=0$. One obtains \eqref{A13} by remarking that, for any complex $z\ne0$, both $z^j e^{(\beta z-\alpha-\beta +\alpha/z)t}$ and $(\beta z/\alpha)^{-j} e^{(\beta z-\alpha-\beta +\alpha/z)t}$ are solutions to \eqref{A10}, and by checking that $\pi_{j,k}(0)=\delta_{j,k}$ and $\pi_{0,k}(t)=0$. The form \eqref{AA24} is obtained by making the change of variable $z=\sqrt{\alpha/\beta}e^{i\theta}$.

We see that equations \eqref{C3}, \eqref{A9}, \eqref{A12}, and \eqref{A13} (or \eqref{AA24}) give \emph{an explicit expression of $u_j(t)$ and $q_t$ at  all times and for an arbitrary initial condition}.

\subsection{Long time asymptotics}\label{sec:saddle}

For fixed $j$ and $k$, one can  extract from \eqref{A13} or \eqref{AA24}
the large $t$ asymptotics of $\pi_{j,k}(t)$ using the saddle point (or Laplace's) method.
When $t$ is large, \eqref{AA24} is dominated by $\theta$ close to~0; more precisely, it is dominated by $\theta$ of order $1/\sqrt t$. 
Making the change of variable $\theta=u/\sqrt t$ and then expanding in powers of $t$ lead to some Gaussian integrals, and we arrive at

\begin{equation}
 \pi_{j,k}(t)    
=
{e^{-(\sqrt{\beta}-\sqrt{\alpha})^2 t}
\over  
\big(t \sqrt{\alpha \beta}\big)^{{3\over 2}} } 
\left({\alpha \over \beta}\right)^{j-k \over 2} 
  {j  k  \over 2 \sqrt{\pi} } 
\left(1+  { 5-4 j^2 - 4 k^2  \over 16  \, \sqrt{\alpha \beta } \, t}  +\mathcal O\Big(\frac1{t^2}\Big) \right) .
\label{A21}
\end{equation}
From \eqref{A12}, we obtain that
\begin{equation}
v_j(t)={e^{-(\sqrt{\beta}-\sqrt{\alpha})^2 t}
\over  
\big(t \sqrt{\alpha \beta}\big)^{{3\over 2}} } 
\left({\alpha \over \beta}\right)^{j\over 2} 
  {j  S  \over 2 \sqrt{\pi} } 
\big(1+  o(1)   \big)  ,
\label{AA30}
\end{equation}
under the condition that the following sum
\begin{equation}
S= 
 \sum_{k \ge 1} k 
\left({\beta \over \alpha}\right)^{k \over 2}
u_k(0) 
\label{A23}
\end{equation}
is finite. (Strictly speaking, in order to write \eqref{AA30}, one needs to check that the $\mathcal O(1/t^2)$ term in~\eqref{A21} does not contribute to the leading term in~\eqref{AA30}. This can be done by dominated convergence, but we omit the argument.)

If, furthermore, the sum $\sum_{k\ge1} k^3 (\frac\beta\alpha)^{\frac k2} u_k(0)$ is also finite, then the $o(1)$ correction in \eqref{AA30} is actually $\mathcal O(1/t)$.

Then, still assuming that $S<\infty$, we compute $\sum_{j\ge1} v_j(t)$ and finally obtain
from \eqref{C3} and \eqref{A9},  that $u_j(t)$ converges as $t\to\infty$ to the maximal $\kappa=\kappa_c$ solution     \eqref{A20} and that
\begin{equation}
q_t= (\sqrt{\beta}-\sqrt{\alpha})^2 t  + {3 \over 2} \ln(t ) +\ln\left[{2 \sqrt{\pi} (\alpha\beta )^{1\over 4}  ( \sqrt{\beta}-\sqrt{\alpha})^2  \over S} \right]+ o(1). 
\label{A22}
\end{equation}

The $\frac32\ln (t)$ correction obtained in \eqref{A22} (assuming that $S$ is finite) is reminiscent of Bramson's shift \cite{Bramson.1978,Bramson.1983} in Fisher-KPP. It is noteworthy that here,   in contrast to the Fisher-KPP problem, the constant term in the large $t$ expansion of $q_t$ can be computed explicitly \emph{for arbitrary initial conditions with $S$ finite}. In addition, if the initial condition decays fast enough, the next term in the large $t$ expansion is $1/t$. This is different from the Fisher-KPP case, where the large $t$ expansion of the position gives universal $1/\sqrt t$ and $\ln(t)/t$ terms before the $1/t$ term~\cite{EbertvanSaarloos.2000,BrunetDerrida.2015,BerestyckiBrunetDerrida.2017,BerestyckiBrunetHarrisRoberts.2016,NolenRoquejoffreRyzhik.2019,Graham.2019}.

When the sum (\ref{A23}) is infinite, \textit{i.e.}\@ when the initial condition does not decay fast enough, the large $t$ asymptotics of $q_t$ are modified. In \autoref{sec:RL0}  where  a continuous space version of the problem is considered, we will discuss in more details   the influence of the initial condition on these  asymptotics. 
Here let us simply consider the case where $u_k(0)$ decays exponentially with~$k$:
\begin{equation}
u_k(0) = b_k+ C  c^{k}
\label{A24}
\end{equation}
where  $\sqrt{\alpha / \beta} < c < 1$ and where $b_k$ decreases fast enough that $\tilde S=\sum_k k (\frac\beta\alpha)^{k/2} b_k$ converges absolutely.
The $b_k$ part gives a contribution to $v_j(t)$ equal to \eqref{AA30} with $S$ replaced by $\tilde S$; this contribution will turn out to be negligible. To compute the contribution of the $C c^k$ term, we take 
in \eqref{A13} a contour where $|z|>c$ everywhere. Then, the sum in \eqref{A12} can be pushed under the integral sign of \eqref{A13} and we obtain
\begin{equation}
v_j(t)=
[\text{contribution of the $b_k$}]+\frac1{2\pi i}\oint_{|z|>c} \frac{\diffd z}z \biggl[z^j - \Bigl(\frac{\beta z}\alpha\Bigr)^{- j}\biggr]\frac{Cc}{z-c}e^{\left(\beta z -\alpha-\beta+\frac\alpha z\right)t}
.
\label{bk}
\end{equation}
In \eqref{bk}, one can replace the contour $|z|> c$ by a circle of radius $\sqrt{\alpha/\beta}$ around the origin and another small circle around the pole $z=c$. The contribution of the first circle and of the $b_k$ are negligible in the long time limit, and $v_j(t)$ is dominated by the residue at $z=c$:
\begin{equation}
v_j(t)\simeq C\biggl[c^j - \Bigl(\frac{\beta c}\alpha\Bigr)^{- j}\biggr]e^{\left(\beta c -\alpha-\beta+\frac\alpha           c\right)t}
.
\label{AA36}
\end{equation}
(The other contributions in \eqref{bk} are of order $e^{-(\sqrt\beta-\sqrt\alpha)^2t}$ as in \eqref{AA30}, and are negligible since
$\beta c -    \alpha-\beta+\frac\alpha           c > - (\sqrt\beta-\sqrt\alpha)^2$ for $c>\sqrt{\alpha/\beta}$.)

From \eqref{AA36}, one finally obtains
\begin{equation}
u_j(t) \to 
{
 (1-c)(\beta c- \alpha)
\over \beta c^2-\alpha 
} 
\left[c^{j} - \Big(\frac{\beta c }\alpha \Big)^{-j} \right],
\end{equation}
which is the steady state solution (\ref{A16}) for $\kappa= \alpha + \beta - {\alpha \over c} - \beta c$,
and  
\begin{equation}
q_t= \left[\alpha + \beta - {\alpha \over c} - \beta c \right] t
-\ln{C(\beta c^2-\alpha) \over (1-c) (\beta c - \alpha)} + o(1).
\end{equation}
Again, as for traveling waves in the Fisher-KPP equation, the asymptotic profile is determined by the way the initial condition decays at infinity.

\section{The transition between the pulled and pushed cases} \label{sec:transition}
In this section we consider a modified version of the Fleming-Viot process of \autoref{sec:lattice} where the rate $ \beta$ is modified on site $1$ to become $\beta_1<\beta$. As $\beta_1$ varies, we are going to see that there is a transition at a critical value $ \beta_1^*= \beta - \sqrt{\alpha \beta}$ very similar to the transition between pulled and pushed fronts for traveling waves~\cite{HadelerRothe.1975,BBDKL.1985,vanSaarloos.2003,GarnierGilettiHamelRoques.2012,AHR.2023,AHR.2023b,AHS.2023,Derrida.2023}.

The evolution \eqref{eq:FV} for the expectation $\langle n_j\rangle$ in the Fleming-Viot process becomes
\begin{equation}
\frac{\diffd  \langle n_j \rangle}{ \diffd  t} = 
\alpha \langle n_{j-1} \rangle +\beta \langle n_{j+1} \rangle 
-(\alpha  +\beta_j )\langle n_{j} \rangle  + \frac{\beta_1}{ N-1} \Big( \langle n_1 n_j \rangle  - \langle n_1 \rangle \delta_{j,1} \Big),
\label{eq:FVpushed}
\end{equation}
where 
\begin{equation}
\beta_j=\begin{cases}\beta_1&\text{for $j=1$},\\
	\beta &\text{for $j>1$}.\end{cases}
\end{equation}
The equation \eqref{A10} for the probability $v_j(t)$ that a single walker jumping with the same rates survives up to time $t$ and sits at position $j$ becomes
\begin{equation}
\frac{\diffd  v_j}{\diffd t} = 
\alpha v_{j-1}  +\beta   v_{j+1} 
-(\alpha  +\beta_j )  v_{j},
\label{A49}
\end{equation}
with $v_0=0$.
We introduce $q_t$ and the probability $u_j(t)$ that the walker is on $j$ at time~$t$ given that it survives,  as  in \autoref{sec:lattice}, see \eqref{C3} and \eqref{A9}:
\begin{equation}
u_j(t)  = e^{q_t}   \, v_j(t), \qquad 1 = e^{q_t}\sum_{j\ge1}v_j(t).
\label{A48}
\end{equation}
Then the evolution equation \eqref{A6} for $u_j(t)$ becomes
\begin{equation}
\frac{\diffd   u_j}{\diffd t} = 
\alpha  u_{j-1}  +\beta   u_{j+1} 
-(\alpha  +\beta_j )  u_{j}  + \beta_1   u_1  u_j,
\label{A47}
\end{equation}
with $u_0=0$.
As in \autoref{sec:lattice}, this is a mean-field version for large~$N$ of \eqref{eq:FVpushed}.

The linear system of equations \eqref{A49} can be exactly solved for any initial condition $v_j(0)=u_j(0)$.
Writing, as in \eqref{A12},
\begin{equation}
v_j(t) = \sum_{k \ge 1} \pi_{j,k}(t)   u_k(0),
\label{AA60}
\end{equation}
we see that $\pi_{j,k}(t)$ is the solution to \eqref{A49} with initial condition $\pi_{j,k}(0)=\delta_{j,k}$.

Adapting the approach which led to \eqref{A13}, the solution is given by
\begin{equation}
\pi_{j,k}(t)=\frac1{2\pi i}\oint \frac{\diffd z}z z^{-k} \varphi_j(z)e^{\left(\beta z -\alpha-\beta+\frac\alpha z\right)t}
\label{AA62}
\end{equation}
with
\begin{equation}
\varphi_j(z)= z^j - \Bigl(\frac{\beta z}\alpha \Bigr)^{1-j}\ {\alpha + (\beta_1-\beta) z \over \beta_1 + \beta(z-1)} \quad\text{for $j\ge1$},\qquad\qquad \varphi_0=0.
\label{AA61}
\end{equation}
where the contour is such that both poles at 0 and $z_1=1-\beta_1/\beta$ are inside.
(Note that, at $t=0$, using a circle of large radius for the contour, one has
 \begin{equation}
 \frac{1}{2\pi i}\times \frac1z z^{-k} \varphi_j(z)=\frac{1}{2\pi i}z^{j-k-1}+\mathcal O(|z|^{-j-k})\quad\text{as $|z|\to\infty$}.
 \end{equation}
The integral of the first term leads to the initial condition $\delta_{j,k}$, while the integral of the second term with $j\ge1$ and $k\ge1$ is zero, since it converges to 0 as the radius of the contour goes to infinity.)
 
In \autoref{sec:solu}, we used as a contour the  circle of radius $z_0=\sqrt{\alpha/\beta}$. Here, we can use the same contour if it contains the pole at $z_1=1-\beta_1/\beta$.
We check that $0<z_1<z_0$ if and only if $\beta_1^*<\beta_1<\beta$ with
\begin{equation}
\beta_1^*=\beta-\sqrt{\alpha\beta}.
\label{AAb1*}
\end{equation}
Then, if $\beta_1^*<\beta_1<\beta$,
\begin{align}
\pi_{j,k}(t)&=\frac1{2\pi i}\oint_{|z|=z_0} \frac{\diffd z}z \varphi_j(z)z^{-k}e^{\left(\beta z -\alpha-\beta+\frac\alpha z\right)t}.
\label{AA64}
\end{align}
On the other hand, if $\beta_1<\beta_1^*$, one must add an extra contour around $z_1$ which leads to
\begin{equation}
\begin{aligned}
\pi_{j,k}(t)&=\frac1{2\pi i}\oint_{|z|=z_0} \frac{\diffd z}z \varphi_j(z)z^{-k}e^{\left(\beta z -\alpha-\beta+\frac\alpha z\right)t}
\\&\qquad+\frac{\beta^{k-1}\alpha^{j-1}}{(\beta-\beta_1)^{j+k}}\left[(\beta_1-\beta)^2-\alpha\beta\right]e^{(\frac{\alpha\beta_1}{\beta-\beta_1}-\beta_1)t}\qquad\text{if  $\beta_1<\beta_1^*$},
\end{aligned}
\label{AA65}
\end{equation}
where the second term  is simply the residue of $\frac1z \varphi_j(z) z^{-k} e^{\left(\beta z -\alpha-\beta+\frac\alpha z\right)t}$ around $z_1$.
In fact, expressions \eqref{AA64} and \eqref{AA65} are analytic continuations of each other when $\beta_1$ crosses the value $\beta_1^*$.

From \eqref{A48}, \eqref{AA60}, \eqref{AA61}, \eqref{AA64} and \eqref{AA65}, one 
can extract the long time asymptotics of $q_t$ and $u_i(t)$. 
The result depends on the sign of $\beta_1-\beta_1^*$
\begin{itemize}
\item
For $\beta_1^*<\beta_1<\beta$ (the pulled case):\\
$\pi_{j,k}(t)$ is given by \eqref{AA64}, which can be evaluated by the saddle point    method at $z=\sqrt{\alpha/\beta}$ as in \autoref{sec:saddle}. One finds that
\begin{equation}
 \pi_{j,k}(t)    
=
{e^{-(\sqrt{\beta}-\sqrt{\alpha})^2 t}
\over  
2\sqrt\pi \big(t \sqrt{\alpha \beta}\big)^{{3\over 2}} } 
\left[{\alpha \over \beta}\right]^{j-k \over 2} 
\biggl[\biggl(j-\frac{\beta_1-\beta}{\beta_1-\beta_1^*}\biggr)
\biggl(k-\frac{\beta_1-\beta}{\beta_1-\beta_1^*}\biggr)+\mathcal O\Big(\frac1t\Big)\biggr].
\label{AA67}
\end{equation}
Introduce $S$ as in (\ref{A23}) and $R$ as follows:
\begin{equation}
R=
 \sum_{k \ge 1} 
\left({\beta \over \alpha}\right)^{k \over 2}
u_k(0), \qquad
S=
 \sum_{k \ge 1} 
\left({\beta \over \alpha}\right)^{k \over 2}
k u_k(0).
\label{A58}
\end{equation}
For initial conditions decaying fast enough for $S$ to be finite, one obtains
\begin{equation}
 v_{j}(t)    
=
{e^{-(\sqrt{\beta}-\sqrt{\alpha})^2 t}
\over  
2\sqrt\pi \big(t \sqrt{\alpha \beta}\big)^{{3\over 2}} } 
\left[{\alpha \over \beta}\right]^{j \over 2} 
\biggl[\biggl(j-\frac{\beta_1-\beta}{\beta_1-\beta_1^*}\biggr)
\biggl(S-\frac{\beta_1-\beta}{\beta_1-\beta_1^*}R\biggr)+ o(1)\biggr].
\end{equation}
where the $o(1)$ is actually a $\mathcal O(1/t)$ if $u_k(0)$ decreases fast enough. Then, by normalization, see \eqref{A48},
\begin{equation}
u_j(t) \to \frac{(\sqrt{\beta}-\sqrt{\alpha})^2}{\alpha\beta_1} \, \Big[( \beta_1  - \beta_1^*) j -\beta_1+ \beta\Big]  \left[\alpha\over \beta\right]^{j \over 2} ,
\label{A56} 
\end{equation}
 and
\begin{equation}
\begin{aligned}
 q_t=  & (\sqrt{\beta}-\sqrt{\alpha})^2 t  + {3 \over 2} \ln(t ) 
+\ln\left[{2 \sqrt{\pi} \beta^{\frac34}   ( \beta_1 - \beta_1^*)^2  (\sqrt{\beta}-\sqrt{\alpha})^2  \over \alpha^{\frac14} \beta_1 [(\beta_1 - \beta_1^*)S + (\beta-\beta_1) R]} \right]+ o(1) .
\end{aligned}
\label{A57}
\end{equation}
The limiting shape in \eqref{A56}  is the steady state solution of (\ref{A47}) with maximal flux $\kappa=\kappa_c=(\sqrt\beta-\sqrt\alpha)^2$.

\item For $\beta_1< \beta_1^*$ (the pushed case):\\
We use (\ref{AA65}) to evaluate $\pi_{j,k}$. The first term is still given by \eqref{AA67}, but it is dominated by the additional term. Assuming that
\begin{equation}
S'= \sum_{k \ge 1} \left({\beta \over \beta-\beta_1} \right)^{k} \, u_k(0)
\label{A61} 
\end{equation}
is finite,
one obtains
\begin{equation}
v_j(t)= S'\frac{\beta^{-1}\alpha^{j-1}}{(\beta-\beta_1)^{j}}\left[(\beta_1-   \beta)^2-\alpha\beta\right]e^{(\frac{\alpha\beta_1}{\beta-\beta_1}-             \beta_1)t}\big(1+o(1)\big)\qquad\text{if  $\beta_1<\beta_1^*$},
\end{equation}
where the corrections are actually exponentially small in time. Using \eqref{A48}, this leads to
\begin{equation}
u_j(t) \to {\beta-\beta_1-\alpha \over \alpha} \left({ \alpha \over \beta-\beta_1 } \right)^j ,
\label{A59} 
\end{equation}
which is the steady state solution of (\ref{A47}) corresponding to the flux $\kappa=\beta_1-\frac{\alpha\beta_1}{\beta-\beta_1}$
and
\begin{equation}
q_t= \bigg(\beta_1-\frac{\alpha\beta_1} {\beta-\beta_1}\biggr) t + \ln\left[{(\beta - \beta_1 - \alpha) \beta \over [(\beta-\beta_1)^2 - \alpha \beta]S'} \right]  +o(1).
\label{A60} 
\end{equation}
(See also \cite{Villemonais.2015}.)
\item For $\beta_1=\beta_1^*+\mathcal O(1/\sqrt{t})$ (the crossover regime):\\
The crossover regime contains the critical case $\beta_1=\beta_1^*$ and connects to the pulled and pushed cases.
Writing
\def\notepsilon{c}
\begin{equation}
\beta_1= \beta_1^*-(\alpha \beta)^{\frac14} { \notepsilon \over \sqrt{t}}  ,
\label{A62} 
\end{equation}
one finds  from \eqref{AA64} (if $\notepsilon<0$) or from \eqref{AA65} (if $\notepsilon>0$) that
\begin{equation}
 \pi_{j,k}(t)    
=
{e^{-(\sqrt{\beta}-\sqrt{\alpha})^2 t}
\over  
\sqrt\pi \big(t \sqrt{\alpha \beta}\big)^{{1\over 2}} } 
\left[{\alpha \over \beta}\right]^{j-k \over 2} 
\Bigl[1+2\notepsilon e^{\notepsilon^2}\int_{-\infty}^\notepsilon e^{-v^2}\,\diffd v+\mathcal O\Big(\frac1{\sqrt t}\Big)\Bigr].
\label{AA50}
\end{equation}
To leading order, assuming $R<\infty$, the large time limit of $u_j(t)$ is  therefore given by 
\begin{equation}
u_j(t)\to\frac{\sqrt\beta-\sqrt\alpha}{\sqrt\alpha}\biggl(\frac\alpha\beta\biggr)^{\frac j 2},
\end{equation}
and $q_t$ is 
\begin{equation}
q_t= (\sqrt{\beta}-\sqrt{\alpha})^2 t + {1\over 2} \ln t  +  \ln {\sqrt{\pi} (\sqrt{\beta}-\sqrt{\alpha}) \beta^{\frac14} \over   \alpha^{\frac14} R}-\Psi_1(\notepsilon) +o(1),
\label{A63} 
\end{equation}
with
\begin{equation}
\Psi_1(c)=\ln\bigg[1+2\notepsilon e^{\notepsilon^2}\int_{-\infty}^\notepsilon e^{-v^2}\,\diffd v\bigg].
\label{A64} 
\end{equation}

If $\notepsilon=(\beta_1^*-\beta_1)\sqrt t\big/{(\alpha\beta)^{\frac14}}$ becomes large, but still smaller than $\sqrt t$, which corresponds to the range $\frac1{\sqrt t}\ll|\beta_1^*-\beta_1|\ll 1$,
one finds that
\begin{equation}\label{AAA81}
\Psi_1(c)\simeq\begin{cases}
c^2+\ln(2c\sqrt\pi)=\frac{(\beta_1^*-\beta_1)^2}{\sqrt{\alpha\beta}}t+\frac12\ln t + \ln\frac{2\sqrt\pi(\beta_1^*-\beta_1)}{(\alpha\beta)^{\frac14}} & \text{if $c\gg1$},\\
-\ln(2c^2)=-\ln t-\ln\frac{2(\beta_1^*-\beta_1)^2}{\sqrt{\alpha\beta}}& \text{if $-c\gg1$}.
\end{cases}
\end{equation}
One can check that (\ref{A63}) with \eqref{AAA81} matches with (\ref{A57}) or (\ref{A60}) (depending on the sign of $\beta_1-\beta_1^*$), meaning that small $\beta_1-\beta_1^*$ expansions of either expression give, to leading order, the same result.
One can notice that the cross-over function $\Psi_1$ is the same as for a traveling wave problem at the transition between pulled and pushed fronts,
see (18) in \cite{Derrida.2023}.

\end{itemize}

\section{The Fleming-Viot process on  the real line }\label{sec:RL0}

On the real half line, the Fleming-Viot process is a set of $N$ independent Brownian motions with drift towards the origin. When a particle touches the origin, it is moved to the position of one of the $N-1$ other particles.

In the hydrodynamic limit, the system is described by a density $u(x,t)$ of particles for $x\ge0$, with $u(0,t)=0$ and $\int u(x,t)\,\diffd x =1$ where
\begin{equation}
\partial_t u = \tfrac12\partial_x^2u + \mu\partial_x u + \tfrac12\partial_x u(0,t)\, u.
\label{B75}
\end{equation}
Indeed, on the right hand side, the first term $\tfrac12\partial_x^2u$ represents diffusion, the second term $\mu\partial_x u$ is the drift towards the origin and the last term is a growth term with a rate $\tfrac12\partial_x u(0,t)$ equal to the flux of particles getting absorbed at the origin.

Let us now consider
a single Brownian motion on the half line with the same drift and with absorption at the origin, and let $v(x,t)\,\diffd x$ be the probability that this Brownian never reached the origin up to time $t$ and is in $\diffd x $ at time $t$. Then
\begin{equation}
\partial_t v = \tfrac12\partial_x^2v + \mu\partial_x v,\qquad \text{with } v(0,t)=0. 
\label{B78}
\end{equation}
Let $e^{-q_t}$ be the probability that the Brownian never reached the origin up to time $t$,
\begin{equation}
e^{-q_t}=\int_0^\infty \diffd x\, v(x,t).
\label{B61}
\end{equation} 
In \eqref{B78}, one obtains that
\begin{equation}
\partial_t q_t = e^{q_t} \frac12\partial_x v(0,t),\label{B62}
\end{equation}
and then one checks easily that the density of probability $e^{q_t} v(x,t)$ that the Brownian is at $x$  at time $t$ \emph{given} it never reached the  origin follows \eqref{B75}. Then, if $u(x,0)=v(x,0)$, one has
\begin{equation}
u(x,t)=e^{q_t} v(x,t).\label{B63}
\end{equation}
From \eqref{B62}, the flux of mass being absorbed is $\partial_t q_t=\frac12\partial_x u(0,t)$, and one obtains that $q_t$ is, as in \autoref{sec:lattice}, the total mass of particles absorbed up to time $t$. (We call ``mass''  a number of particles divided by $N$, in the $N\to\infty$ limit.)

The equations \eqref{B75}, \eqref{B78}, \eqref{B61} and \eqref{B63}
should be respectively compared to the equations \eqref{eq:FV} (or \eqref{A6}), \eqref{A10}, \eqref{C3} and \eqref{A9} of \autoref{sec:lattice}.
In fact, one can  obtain the continuous space version of this section from the lattice version of  \autoref{sec:lattice} by  choosing  $\alpha$ and $\beta$ of the form
\begin{equation}
\alpha=\frac{1-\mu\epsilon}2,\qquad
\beta =\frac{1+\mu\epsilon}2.
\label{eq:scaling1}
\end{equation}
Then, by rescaling space and time, we relate $v_j(t)$ and $u_j(t)$ to the new functions $u(x,t)$ and $v(x,t)$ in the following way:
\begin{equation}
v_j(t) = \epsilon v( j\epsilon, t\epsilon^2),\qquad
u_j(t) = \epsilon u( j\epsilon, t\epsilon^2),
\label{eq:scaling}
\end{equation}
we obtain \eqref{B75} and \eqref{B78} from  \eqref{A6} and \eqref{A10} in the $\epsilon\to0$ limit.

\subsection{Steady states solutions}
In a steady state solution of \eqref{B75}, the mass of particles is absorbed at a constant rate
\begin{equation}
q_t=\kappa t\qquad\text{with } \kappa =\tfrac12\partial_x u(0,t),
\end{equation}
and $u$ is solution to $\frac12\partial_x^2u + \mu \partial_x u + \kappa u=0$ with $u(0,t)=0$.
One finds easily that the solutions are
\begin{align}
u(x)&=\frac\kappa{\sqrt{\mu^2-2\kappa}}\bigg[e^{-\left(\mu-\sqrt{\mu^2-2\kappa}\,\right)x} - e^{-\left(\mu+\sqrt{\mu^2-2\kappa}\,\right)x}\bigg] &&\text{if $0<\kappa<\kappa_c=\frac{\mu^2}2$},
\label{noncritcont}
\\
u(x)&=\mu^2 x e^{-\mu x} && \text{if $\kappa=\kappa_c$}. \label{critcont}
\end{align}
There are no positive solution if $\kappa\le0$ or $\kappa>\kappa_c$.

\subsection{Long time asymptotics for arbitrary initial conditions}

As in  \autoref{sec:lattice}, we write the solution to the linear equation \eqref{B78} as
\begin{equation}
v(x,t)=\int_0^\infty \diffd y \, \pi(x,y,t) u(y,0),
\label{vpu0}
\end{equation}
where $\pi(x,y,t)$ is the transition probability density of the diffusion with drift absorbed at 0 from $y$ to $x$, meaning that $\pi(x,y,t)\,\diffd x$ is the probability that the diffusion, started from $y$, does not get absorbed up to time $t$ and is in $\diffd x$ at time $t$. It is the solution to \eqref{B78} with initial condition $\pi(x,y,0)=\delta(x-y)$. The method of mirrors gives, classically,
\begin{equation}
\pi(x,y,t)= \frac1{\sqrt{2\pi t}}\left[e^{-\frac{(x+\mu t-y)^2}{2t}}- e^{2\mu y-\frac{(x+\mu t+y)^2}{2t}}\right]=\sqrt{\frac2{\pi t}}e^{-\frac{\mu^2}2t-\mu x-\frac{x^2}{2t}}\sinh\frac{xy}t e^{\mu y -\frac{y^2}{2t}}.
\label{AA80}
\end{equation}
This expression can also be obtained by using the scaling \eqref{eq:scaling} with the expression \eqref{AA24} of $\pi_{i,j}(t)$.

One can extract the large time asymptotics of $v(x,t)$, $u(x,t)$ and $q_t$ from the explicit expressions \eqref{vpu0}, \eqref{AA80}, \eqref{B61} and \eqref{B63},  and analyze how they depend on the decay of the initial condition $u(y,0)$.
When $t$ is large, for $x=\mathcal O(\sqrt t)$ and $y$ of order 1, one gets from~\eqref{AA80}
\begin{equation}
\pi(x,y,t)= \sqrt{\frac2{\pi t^3}}x e^{-\frac{\mu^2}2t-\mu x-\frac{x^2}{2t}}e^{\mu y}\bigg(y+\frac{x^2y^3}{6t^2}-\frac{y^3}{2t}+o\Big(\frac1t\Big)\bigg).
\label{C71}
\end{equation}
Let
\begin{equation}
Y_p = \int_0^\infty \diffd y\, y^p e^{\mu y} u(y,0).
\end{equation}
\begin{itemize}
\item   If  $u(y,0) $ decays fast enough   that $Y_1<\infty$, one obtains from \eqref{vpu0} and \eqref{C71}
\begin{equation}
v(x,t)= 
\sqrt{\frac2{\pi t^3}}x e^{-\frac{\mu^2}2t-\mu x-\frac{x^2}{2t}}\big(Y_1+ o(1)\big) .
\label{A73}
\end{equation}
If, furthermore, $Y_3<\infty$, the $o(1)$ is in fact $Y_3\big(\frac{x^2}{6t^2}-\frac1{2t}\big)+o\big(\frac1t\big)$. (Recall that we assume $x=\mathcal O\big(\sqrt t\big)$.)
Integrating \eqref{A73} over $x$ to obtain $e^{-q_t}$ leads to
\begin{equation}
q_t=\frac{\mu^2}2t +\frac32\ln t + \ln \frac{\mu^2\sqrt\pi}{Y_1\sqrt 2} +o(1),
\label{C75}
\end{equation}
and, still with $x=\mathcal O(\sqrt t)$,
\begin{equation}
u(x,t)= \mu^2 x e^{-\mu x-\frac{x^2}{2t}}\big(1+o(1)\big).
\label{A38}
\end{equation}
In particular, $u(x,t)$ converges to the critical steady state solution \eqref{critcont} if $x=o\big(\sqrt t\big)$. If $Y_3<\infty$, the $o(1)$ in \eqref{C75} is in fact $\frac1{2t}\big(\frac6{\mu^2}+\frac{Y_3}{Y_1}\big)+o\big(\frac1t\big)$.
There is no $1/\sqrt t$ term for initial conditions that decay fast enough.

\item If,  for large  $y$,
\begin{equation}
u(y,0) \simeq A  y^\nu  e^{-\mu y}\qquad\text{with }\nu>-2,
\end{equation}
then from \eqref{AA80}
\begin{equation}
\begin{aligned}
v(x,t)&\simeq A
\sqrt{\frac2{\pi t}}e^{-\frac{\mu^2}2t-\mu x-\frac{x^2}{2t}}\int_0^\infty \diffd y\, y^\nu\sinh\frac{xy}t e^{ -\frac{y^2}{2t}}.
\\&\simeq A
\sqrt{\frac2{\pi t^3}} x e^{-\frac{\mu^2}2t-\mu x}\int_0^\infty \diffd y\, y^{\nu+1} e^{ -\frac{y^2}{2t}}&&\text{if $x=\mathcal O(1)$}
\\&\simeq A
\sqrt{\frac2{\pi t^3}} x e^{-\frac{\mu^2}2t-\mu x}2^{\frac\nu2}t^{1+\frac\nu2}\Gamma\Big[1+\frac\nu2\Big]&&\text{if $x=\mathcal O(1)$}
\end{aligned}
\end{equation}

One gets in the long time limit
\begin{equation}
q_t\simeq \frac{\mu^2}{2}t +\frac{1-\nu}2\ln t + \ln\bigg[\frac{\mu^2\sqrt\pi}{A\,2^{\frac{\nu+1}2}\Gamma\big(1+\frac\nu2\big)}\bigg],
\end{equation}
and for $x=\mathcal O\big(\sqrt{t}\big)$,
\begin{equation}
u(x,t)\simeq \mu^2 e^{-\mu x} \sqrt{\frac t2} H_\nu\bigg(\frac{x\sqrt 2}{\sqrt t}\bigg)
\label{A42}
\end{equation}
with
\begin{equation}
H_\nu(z) =\frac{2 e^{-\frac{z^2}{ 4} }}{\Gamma\big(1+\frac\nu 2\big) } \int_0^\infty\diffd w\, w^\nu  e^{-w^2} \ \sinh( z w ).
\end{equation}
$H_\nu$ is an hypergeometric function which satisfies. 
\begin{equation}
2 H_\nu'' + x H_\nu' - \nu H_\nu=0,\qquad H_\nu(0)=0,\qquad H_\nu'(0)=1
\label{A43}
\end{equation}
If $x=o\big(\sqrt t\big)$, then \eqref{A42} is the critical steady state \eqref{critcont}.
 Note that $H_\nu(z) \simeq  2 \sqrt{\pi} z^\nu /\Gamma(1+\frac\nu 2)$ for large $z$,  and that for some values of $\nu$ it takes simpler expressions (for example  $H_1 = z$, $H_3 = z + z^3/6$).
The scaling function $H_\nu$ is exactly the same as the  corresponding scaling function  in the Fisher-KPP equation: $H_\nu= G_{1-\frac\nu 2}$ of~\cite{BrunetDerrida.1997}.

\item
If, for large $y$
\begin{equation}
u(y,0) \simeq A e^{-\gamma y}, \qquad\text{with } \gamma<\mu,
\end{equation}
one gets  for large $t$
\begin{equation}
v(x,t)\simeq A e^{-\frac{2\mu-\gamma}2\gamma t}\Big[e^{-\gamma x}-e^{-(2\mu-\gamma)x}\Big].
\end{equation}
and
\begin{equation}
q_t=\frac{2\mu-\gamma}2\gamma t +\ln\bigg[\frac{\gamma(2\mu-\gamma)}{2A(\mu-\gamma)}\bigg],\qquad
u(x,t)=\frac{\gamma(2\mu-\gamma)}{2(\mu-\gamma)}\Big[e^{-\gamma x}-e^{-(2\mu-\gamma)x}\Big].
\end{equation}
One recovers the non-critical steady state \eqref{noncritcont} corresponding to $\kappa=\frac{2\mu-\gamma}2\gamma$.
\end{itemize}
In all cases, the various asymptotics are very similar to the ones in Fisher-KPP~\cite{Bramson.1983}.

\section{Pulled-pushed transition on the real line}\label{sec:RL1}

On the lattice, we obtained an analog of the pushed-pulled transition of travelling waves by lowering the rate at which particles jump from 1 to 0. The dynamics remained unchanged elsewhere. We now write what the lattice equations \eqref{A47} become in the continuum.

The hydrodynamics equation \eqref{B75} for $x>0$ remains valid, except that the growth rate in the last term is no longer given by $\frac12\partial_x u(0,t)$. In a generic way, we write
\begin{equation}
\partial_t u = \tfrac12\partial_x^2u + \mu\partial_x u + \partial_t q_t\, u\text{\quad for $x>0$},
\label{PP1}
\end{equation}
with as usual $q_t$ the total mass absorbed up to time $t$.

The effect of $\beta_1<\beta$ on the lattice is to slow down the absorption of particles at the origin. In the continuum version, this leads to having a non-zero density of particles at the origin. Then,  the absorption rate is proportional to that density:
\begin{equation}
\partial_t q_t = a u(0,t)\qquad \text{with }u(0,t)>0,
\label{PP2}
\end{equation}
where $a>0$ is some parameter.
Imposing normalization, $\int_0^\infty \diffd x\,u(x,t)=1$ we obtain from~\eqref{PP1} the following boundary condition:
\begin{equation}
\frac12 \partial_x u(0,t) = (a-\mu) u(0,t).
\label{PP3}
\end{equation}
Equations \eqref{PP1}, \eqref{PP2}, \eqref{PP3} define our model in the continuum. Writing as usual $u(x,t)=e^{q_t}v(x,t)$, we obtain
\begin{equation}
\partial_t v =\frac12\partial_ x^2 v +\mu \partial_x v\text{\quad for $x>0$},\qquad
\frac12 \partial_x v(0,t) = (a-\mu) v(0,t).
\label{PP4}
\end{equation}

The boundary condition in \eqref{PP4} can be obtained from \eqref{A49} for $j=1$, \textit{i.e.} $\partial_t v_1=\beta v_2-(\alpha+\beta_1)v_1$ in the scaling limit \eqref{eq:scaling}, using \eqref{eq:scaling1} and $\beta_1=a\epsilon$. 

Writing as usual
\begin{equation}
v(x,t)=\int_0^\infty \diffd y \, \pi(x,y,t) u(y,0),
\end{equation}
one finds that the solution to \eqref{PP3}, \eqref{PP4} is obtained with
\begin{equation}
\pi(x,y,t)=
 \frac1{\sqrt{2\pi t}}e^{-\frac{\mu^2}2t-\mu(x-y)-\frac{x^2+                      y^2}{2t}}\biggl[e^{\frac{xy}{t}}
+ e^{-\frac{xy}{t}}
-2be^{-\frac{xy}{t}}\int_0^\infty\diffd z\,e^{-(b+\frac{x+y}t)z-\frac {z^2}{2t}}
\biggr]
\label{PP6}
\end{equation}
where 
\begin{equation}
b=2a-\mu.
\end{equation}
(Note that $b=2(\beta_1-\beta_1^*)/\epsilon$ in the scaling limit.)
One can check directly that $\pi(x,y,t)$ satisfies \eqref{PP3}, \eqref{PP4} and $\pi(x,y,0)=\delta(x-y)$; one can also obtain \eqref{PP6} as the scaling limit  of \eqref{AA62}. 
Appendix~\ref{app:Bdar} gives an alternative derivation of \eqref{PP6}, by interpreting $v(x,t)$ as the density of probability of a Brownian motion with drift $-\mu$ (towards the origin since $\mu>0$), reflected at the origin, and with absorption depending on the local time spent at the origin.

As $t\to\infty$, the value of the integral in \eqref{PP6} depends strongly on the sign of $b$. We obtain, for $x$ and $y$ of order~1:
\begin{itemize}
\item If $b>0$, then the integral is dominated by $z$ of order 1 and
one obtains
\begin{equation}
\pi(x,y,t)= \frac2{\sqrt{2\pi}\,t^{3/2}}e^{-\frac{\mu^2}2t-\mu(x-y)}\bigg[xy+\frac{x+y}b+\frac1{b^2}+ \mathcal O(1/t)\biggr].
\end{equation}
which is the scaling limit of \eqref{AA67}.
\item If $b=0$, then
\begin{equation}
\pi(x,y,t)= \frac2{\sqrt{2\pi t}}e^{-\frac{\mu^2}2t-\mu(x-y)}\Bigl[1+ \mathcal O(1/t)\Bigr].
\end{equation}
which is the scaling limit of \eqref{AA50} with $c=0$.
\item If $b<0$, then, the integral is dominated by $z=-bt-x-y$. A saddle point method gives, up to exponentially small terms 
\begin{equation}
\pi(x,y,t)\simeq -2b e^{\frac{b^2-\mu^2}2t+(b-\mu)x+(b+\mu)y}.
\end{equation}
\end{itemize}
One could then recover directly in the continuum all the results presented in \autoref{sec:transition}.

\section{Correction due to a cut-off}\label{sec:cut-off}

So far, in this work, all our results concerned the Fleming-Viot problem in the $N \to \infty$ limit.
For large but finite $N$, the densities of particles and the rescaled number of absorbed particles $Q_t/N$ fluctuate.
Trying to quantify these finite size effects is a difficult problem, of the same nature as understanding the effect of noise in the noisy Fisher-KPP equation or in the $N$-BBM. 

In this section, we adapt to the Fleming-Viot problem the cut-off approximation \cite{BrunetDerrida.1997}, which is known to give the correct leading large $N$ correction in $(\ln N)^{-2}$ to the velocity of noisy Fisher-KPP fronts or of the $N$-BBM in the pulled case~\cite{MuellerMytnikQuastel.2008,BerardGouere.2010}. Whether these cut-off corrections match the large~$N$ corrections of the Fleming-Viot problem is still an open problem.

The cut-off approximation consists in forcing the density $u_j(t)$ on site $j$ to be 0 whenever its value is smaller than a given cut-off value of order~$1/N$. This is intended to match the behaviour of a particle system, where $u_j(t)$ represents the fraction of particles at site $j$ and, therefore, cannot have a non-zero value smaller than~$1/N$. 
Note that the cut-off theory is a deterministic approximation~\cite{BrunetDerrida.1997,KesslerNerSander.1998,DumortierPopovircKaper.2007}, leading to a deterministic flux $\kappa$ of particles.
Here we determine the steady-state of the system in this cut-off approximation.

We will obtain corrections  of order $(\ln N)^{-2}$ in the pulled case and of order $N^{-\gamma}$ with $0<\gamma<1$ in the pushed case, with the same crossover function in the critical regime as in the case of travelling waves~\cite{Derrida.2023}.

When $N=\infty$, the Fleming-Viot problem has infinitely many steady states; the densities of particles in these steady states are the solutions to \eqref{A47} with  $\partial_t u_j=0$:
\begin{equation}
\alpha  u_{j-1}  +\beta   u_{j+1} 
-(\alpha  +\beta_j )  u_{j}  = -\kappa u_j,
\qquad\sum_j u_j=1,\qquad\text{with }\kappa= \beta_1 u_1,
\label{92}
\end{equation}
where $\kappa$ is the rate at which particles are absorbed. Recall that $\beta_j=\beta_1$ for $j=1$ and $\beta_j=\beta$ for $j>1$.
In~\autoref{sec:transition}, we have seen that the $N=\infty$ critical steady state, the one with a maximal flux, 
is given by \eqref{A56} and \eqref{A59}:
\begin{align}
u_j &= C 
\Big[( \beta_1  - \beta_1^*) j -\beta_1+        \beta\Big] z_c^j 
&\kappa&=\kappa_c
&&\text{if $\beta_1>\beta_1^*$ (pulled case),}
\label{ujpulled}
\\
u_j &= \tilde C z_1^j
&\kappa&=\kappa_1
 &&\text{if $\beta_1<\beta_1^*$ (pushed case),}
\label{ujpushed}
\end{align}
with
\begin{equation}\begin{aligned}
\beta_1^* &= \beta-\sqrt{\alpha\beta},&
z_c&=\sqrt{\frac\alpha\beta},&
\kappa_c&=\alpha+\beta-2\sqrt{\alpha\beta}, &
C&=\frac{(\sqrt{\beta}-\sqrt{\alpha})^2}{\alpha\beta_1},\\
&&z_1&=\frac\alpha{\beta-\beta_1},&
\kappa_1&=\beta_1-\frac{\alpha\beta_1}{\beta-\beta_1},&
 \tilde C &=\frac{\beta-\beta_1-\alpha } \alpha.
\end{aligned}
\label{95}
\end{equation}
(See \eqref{AAb1*}, \eqref{A56}, \eqref{A57}, \eqref{A59}, \eqref{A60}. Both expressions \eqref{ujpulled} and \eqref{ujpushed} coincide at $\beta_1=\beta_1^*$.)

For large but finite $N$, the  steady state of the cut-off approximation is close to either \eqref{ujpulled} or \eqref{ujpushed}, depending on $\beta_1$. However, when $N$ is finite, there is always a rightmost non-zero site, which is typically at the  position $L$  where \eqref{ujpulled} or \eqref{ujpushed} reaches~$1/N$:
\begin{equation}
L=\begin{cases}
\displaystyle
\frac{\ln N}{-\ln z_c}+o(\ln N)=\frac{2\ln N}{\ln(\beta/\alpha)}+o(\ln N)&\text{if $\beta_1>\beta_1^*$ (pulled case),}\\[3ex]
\displaystyle
\frac{\ln N}{-\ln z_1}+o(\ln N)=\frac{\ln N}{\ln \frac{\beta-\beta_1}\alpha}+o(\ln N)&\text{if $\beta_1<\beta_1^*$ (pushed case),}
\end{cases}
\label{L96}
\end{equation}

The steady state front in the cut-off approximation must therefore be a solution of~\eqref{92} which vanishes at position~$j=L$.
(We did not consider these steady states earlier in \autoref{sec:transition} when we had $N=\infty$ because they change sign.)

Apart from \eqref{ujpulled}, all the solutions to \eqref{92} are of the form
\begin{equation}
u_j = A(z) z^j + B(z) \Big(\frac\alpha{\beta z}\Big)^j\quad\text{for $j\ge1$}\qquad\text{with }\kappa=\alpha+\beta-\frac\alpha z -\beta z,
\label{93}
\end{equation}
with $\frac\alpha\beta<|z|<1$ (so that $u_j\to0$ as $j\to\infty$) and where $A(z)$ and $B(z)$, which also depend on $\alpha$, $\beta$ and $\beta_1$, are such that $u_j$ is normalized and the equation \eqref{92} is satisfied for $j=1$.
For instance when $z=z_1$, one can check that $B(z_1)=0$ and one recovers \eqref{ujpushed}.

Let us now derive the correction due to a cut-off $1/N$  by identifying
the steady states, \textit{i.e.}\@ solutions to \eqref{92}, which vanish at position $L$ given by \eqref{L96} in the pulled case, the pushed case, the critical case, and in the whole critical regime.

\subsection{The pulled case}
In the pulled case, one has
$\beta>\beta_1>\beta_1^*$.
The steady state \eqref{ujpulled} can be obtained from \eqref{93} by taking the limit $z\to z_c$. Writing $z=z_ce^\epsilon$ and taking $\epsilon\to0$ in \eqref{93}, we see that in order to recover \eqref{ujpulled} one must have
\begin{equation}
A(z_ce^\epsilon)+B(z_ce^\epsilon)\xrightarrow[\epsilon\to0]{} C(\beta-\beta_1),\qquad
\epsilon\big[A(z_ce^\epsilon)-B(z_ce^\epsilon)\big] \xrightarrow[\epsilon\to0]{} C(\beta_1-\beta_1^*).
\end{equation}
When $0<\epsilon\ll1$ we thus obtain that the steady state for $z=z_ce^\epsilon$ is approximatively equal to
\begin{equation}
\begin{aligned}
u_j&= A(z_ce^\epsilon) z_c^j e^{\epsilon j}+B(z_ce^\epsilon) z_c^j e^{-\epsilon j}
\\&\simeq C\biggl[(\beta-\beta_1)\frac{e^{\epsilon j}+e^{-\epsilon j}}2+
(\beta_1-\beta_1^*) \frac{e^{\epsilon j}-e^{-\epsilon j}}{2\epsilon}\biggr]z_c^j
\end{aligned}
\label{102}
\end{equation}
($\epsilon$ is assumed to be small, but $\epsilon j$ might not be.)

This does not vanish when $\epsilon$ is real, but if one takes $\epsilon=\frac{i\pi}\ell$ (with $\ell\gg1$), one obtains
\begin{equation}
u_j= C\Big[(\beta_1-\beta_1^*)\tfrac \ell\pi\sin\tfrac{\pi j}\ell + (\beta-\beta_1) \cos\tfrac{\pi j}\ell\Big]
z_\text c^j,
\label{ujell}
\end{equation}
which vanishes at $j\simeq \ell -\frac{\beta-\beta_1}{\beta_1-\beta_1^*}$. The corresponding absorption rate is equal to
\begin{equation}
\kappa=\alpha+\beta-\frac{\alpha}{z_ce^{\frac{i\pi}\ell}}- \beta z_ce^{\frac{i\pi}\ell} = \alpha+\beta-2\sqrt{\alpha\beta}\;\cos\frac\pi \ell
\simeq\kappa_c+\sqrt{\alpha\beta}\frac{\pi^2}{\ell^2}.
\label{104}
\end{equation}
Writing to leading order that $L\simeq \ell$ and using $L\simeq(\ln N)/|\ln z_c|$, see \eqref{L96}, we finally obtain
\begin{equation}
\label{pul}
\kappa-\kappa_c \simeq \sqrt{\alpha\beta}\,\frac{\pi^2}{L^2}
\simeq \frac{\pi^2\sqrt{\alpha\beta} \ln(\beta/\alpha)^2}{4(\ln N)^2}\qquad\text{(pulled case)}.
\end{equation}
One can notice that neither $\kappa_c$ nor the leading order correction depend on $\beta_1$.

\subsection{The critical case}
When $\beta_1=\beta_1^*$, \eqref{ujell} becomes
\begin{equation}
u_j= C\sqrt{\alpha\beta} \cos\tfrac{\pi j}\ell.
\end{equation}
This steady state vanishes at $j =\ell/2$. Therefore, one must take $\ell=2L$ and one finds that the correction to the absorption rate is four times smaller than in the pulled case:
\begin{equation}
\label{cri}
\kappa-\kappa_c \simeq \sqrt{\alpha\beta}\,\frac{\pi^2}{4L^2}
\simeq \frac{\pi^2\sqrt{\alpha\beta} \ln(\beta/\alpha)^2}{16(\ln N)^2}\qquad\text{(critical case)}.
\end{equation}

\subsection{The pushed case}
In the pushed case, we have
\begin{equation}
\beta>\beta_1^*>\beta_1.
\label{108}
\end{equation}
We look for a solution close to \eqref{ujpushed}. Using $z=z_1$ in \eqref{93} and comparing to \eqref{ujpushed}, we see that $B(z_1)=0$. Using now $z=z_1e^\epsilon$ in \eqref{93} with $\epsilon\ll1$, we obtain
\begin{equation}
\begin{aligned}
u_j&=A(z_1e^\epsilon)(z_1e^\epsilon)^j + B(z_1e^\epsilon) 
\Big(\frac{\alpha e^{-\epsilon}}{\beta z_1}\Big)^j
\\&\simeq A(z_1)\bigg[ z_1^j +z_1\epsilon \frac{B'(z_1)}{A(z_1)}\Big(\frac\alpha{\beta z_1}\Big)^j\bigg] .
\end{aligned}
\label{ujp}
\end{equation}
We used, with a bit of anticipation,  $\epsilon j\ll1$; as we will see now $\epsilon j$ is at most of order $N^{-\gamma} \ln N$ for some $\gamma>0$.

We have already implicitly computed $B(z)/A(z)$ in \eqref{AA61}:
\begin{equation}
\frac{B(z)}{A(z)}=-
\frac{\beta z}\alpha\  \frac{\alpha + (\beta_1- \beta) z }{ \beta_1 + \beta(z-1)}.
\end{equation}
From this with \eqref{95}, we obtain
\begin{equation}\begin{aligned}
\frac{B'(z_1)}{A(z_1)}&=-
\frac{\beta z_1}\alpha\  \frac{\beta_1- \beta}{\beta_1 + \beta(z_1-1)}
=\frac\beta{\beta z_1 - \frac\alpha{z_1}}
.\end{aligned}
\end{equation}
Writing that the steady state \eqref{ujp} vanishes when $j=L=\frac{\ln N}{-\ln z_1}$, we obtain
\begin{equation}
\epsilon = \Big(\frac{\alpha}{\beta z_1^2}-1\Big)\Big(\frac{\beta z_1 ^2}\alpha\Big)^\frac{\ln N}{-\ln z_1}= \Big(\frac{\alpha}{\beta z_1^2}-1\Big)N^{-\gamma}\quad\text{with }\gamma=\frac{\ln\frac{\beta z_1^2}\alpha}{\ln z_1}.
\label{110}
\end{equation}
One checks that $0<\gamma<1$.
The absorption rate is then
\begin{equation}
\kappa = \alpha+\beta-\frac{\alpha}{z_1e^\epsilon}-\beta z_1e^\epsilon
\simeq\kappa_1+\Big(\frac\alpha{z_1}-\beta z_1\Big)\epsilon.
\label{111}
\end{equation}
Finally,
\begin{equation}
\kappa-\kappa_1\sim N^{-\gamma}\qquad\text{(pushed case)}
\label{pus}
\end{equation}
where $\gamma$ is given in \eqref{110}.  The correction is much smaller than in the pulled or critical cases~\cite{KesslerNerSander.1998,BenguriaDepassierHaikala.2007,BenguriaDepassier.2007}.
Note that substituting~\eqref{110} into~\eqref{111} leads to \eqref{pus} with a definite prefactor,
but the value of that prefactor would depend on $L$ being exactly equal to $\ln N/(-\ln z_1)$. If we have, instead, $L=\ln N/(-\ln z_1)+\text{Cste}$, we would obtain the same exponent as in \eqref{pus} with a different prefactor. This is in contrast with  the pulled regime, where adding a constant to $L$ does not change the prefactor in~\eqref{pul}.


\subsection{The critical regime}
We now consider the crossover regime where 
\begin{equation}
\beta_1= \beta_1^* - \frac  {c} L\sqrt{\alpha\beta}
\end{equation}
for some constant $c$,
with $L=\ln N /(-\ln z_c)$ as in \eqref{L96}. (In this regime, one has $z_1\simeq z_c$.) We rewrite the steady state \eqref{ujell} for some value of $\ell$ to be determined using the quantities $x$ and $\chi$ defined by
\begin{equation}
j=xL,\qquad \chi=\frac{\pi L}\ell.
\end{equation}
then, using $\beta-\beta_1\simeq \beta-\beta_1^*=\sqrt{\alpha\beta}$, we obtain
\begin{equation}
{u_{j=xL}}={C\sqrt{\alpha\beta}}\bigg[-\frac c \chi \sin (x \chi) +\cos(x \chi)\bigg]z_c^{xL}.
\label{115}
\end{equation}
For each value of $c$, we must choose $\chi$ in such a way that \eqref{115} is positive for $0\le x < 1$ and zero for $x=1$. This means that $\chi$ must be solution to
\begin{equation}
\chi= c \tan \chi
\label{cotan}
\end{equation}
Then, knowing $\chi$, we obtain $\ell$ and the correction to the absorbing rate using \eqref{104}.
\begin{equation}
\kappa-\kappa_c\simeq\sqrt{\alpha\beta}\,\frac{\pi^2}{\ell^2}
      \simeq\sqrt{\alpha\beta}\,\frac{\chi^2}{L^2}.
\label{121}
\end{equation}

When $c\to-\infty$, we have $\chi\to\pi^-$ and we recover the correction \eqref{pul} of the pulled case.

When $c=0$, we have $\chi=\frac\pi2$ and we recover the correction \eqref{cri} of the critical case.

As $c$ increases, $\chi$ decreases, until the point where $c=1$ and $\chi=0$. 
At this point, the front is $u_{j=xL}=C\sqrt{\alpha\beta}\,(1-x)$ and the leading correction to the absorbing rate is zero ($\kappa=\kappa_c$).

For $c>1$, we need to take $\chi$ a pure imaginary number, and so $\ell$ is also an imaginary number. 
Writing $\chi=i\tilde \chi$ with $\tilde \chi$ a real number in \eqref{115} and \eqref{cotan}, we obtain
\begin{equation}
{u_{j=xL}}={C \sqrt{\alpha\beta}}\bigg[-\frac c {\tilde \chi} \sinh (x \tilde \chi) +\cosh(x \tilde \chi)\bigg]z_c^{xL},\qquad \tilde \chi = c \tanh \tilde \chi.
\label{115x}
\end{equation}

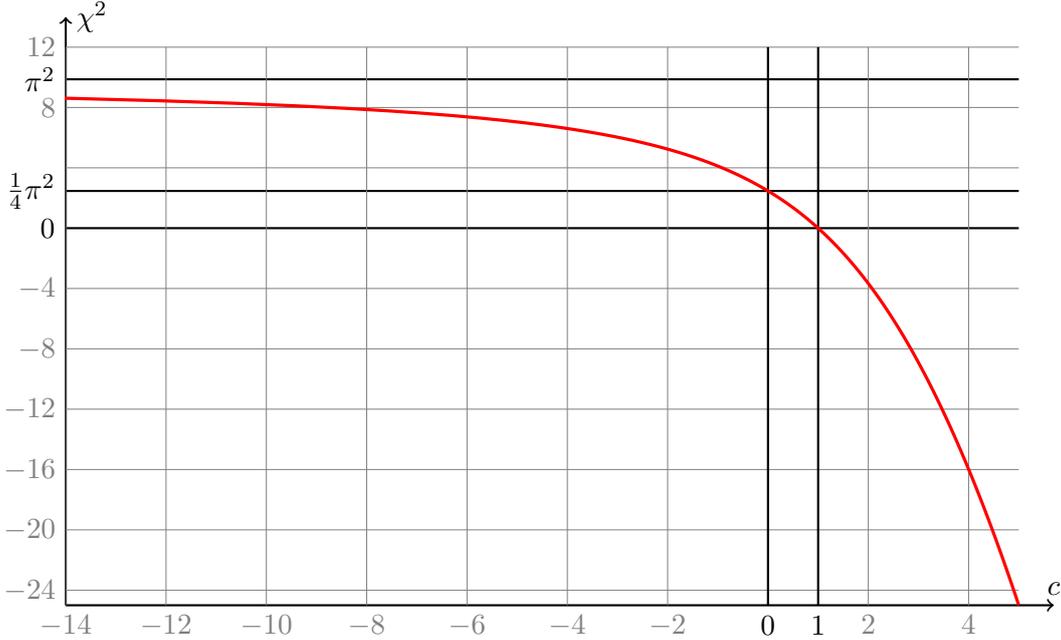
\begin{figure}[ht]\centering
\begin{tikzpicture}[xscale=.66,yscale=.2]
\def\xmax{14}
\def\xmin{-5}
\def\ymin{-25}
\draw[thick,->] (\xmin,\ymin) -- ({\xmax+.7},\ymin) node[above]{$c$};
\draw[thick,->] (\xmin,\ymin) -- (\xmin,14) node[right]{$\chi^2$};
\foreach \y in {-4,-8,...,\ymin} \draw[very thin, black!50] (\xmin,\y) node[left]{$\y$}--++({\xmax-\xmin},0);
\foreach \y/\n in {4/, 8/8,12/12} \draw[very thin, black!50] (\xmin,\y) node[left]{$\n$}--++({\xmax-\xmin},0);
\foreach \y/\n in {0/0, {pi^2/4}/\tfrac14\pi^2,{pi^2}/\pi^2}
	    \draw[thick] (\xmin,\y) node[left]{$\n$}--++({\xmax-\xmin},0);
\foreach \x in {-14,-12,...,-2,2,4} \draw[very thin, black!50]  (9+\x,\ymin) node[below]{$\x$} --++(0,{12-\ymin});
\foreach \x in {0,1} \draw[thick] (9+\x,\ymin) node[below]{$\x$} --++(0,{12-\ymin});
\draw[very thick,red]
 plot[domain=-5:-.05,samples=100] ({9+\x/tanh(\x)},{\x^2}) --
 plot[domain=0.05:2.936,samples=100] ({9+\x/tan(\x/3.1416*180)},{\x^2});
\end{tikzpicture}
\caption{$\chi^2$ as a function of $c$. The correction to the absorption rate is then given by~\eqref{121}.}
\label{fig1}
\end{figure}

\autoref{fig1} shows $\chi^2=-\tilde \chi^2$ as a function of $c$.

\section{Conclusion}

In the $N\to\infty$ limit, the density of particles in the Fleming-Viot problem on positive integers satisfies \eqref{A6}. In this paper, we have obtained exact expressions for this density at finite $t$ for any initial condition, see \autoref{sec:solu}, and we have studied their long time asymptotics. Letting $q_t=\lim_{N\to\infty}\langle Q_t\rangle/N$ where $Q_t$ is the number of particles absorbed up to time $t$ in the $N$ particle system, we have found that  $q_t$ and the position of the front in travelling wave problems varies with time and with the initial condition in very similar ways; in particular, the Fleming-Viot problem has a $\frac32\ln t$ correction to $q_t$ for initial conditions that decay fast enough, similarly to Bramson's shift in the position in travelling fronts. However, there are no $1/\sqrt t$ or $(\ln t)/t$ higher order correction as in travelling waves.

By modifying the rate $\beta_1$ at which particles jump from the leftmost site to the absorbing site, one can still obtain exact expressions for the density and one observes a transition as $\beta_1$ crosses a critical value $\beta_1^*$, see \autoref{sec:transition}. Here again, the long time behaviour of $q_t$ is very similar to the long time behaviour of the position of fronts undergoing a pulled-pushed transition.

All our results can be adapted to Fleming-Viot problem in the continuum, on the half-line, see sections~\ref{sec:RL0} and~\ref{sec:RL1}. In particular, the pulled-pushed transition is obtained by changing the way a Brownian motion is absorbed at the origin.

In \autoref{sec:cut-off}, we have determined the steady state of the Fleming-Viot system when a cut-off~$1/N$ is added to the hydrodynamic equation and have shown that the correction to the flux $\kappa$ of absorbed particles
is of order $(\ln N)^{-2}$ in the ``pulled'' case and $N^{-\gamma}$ for some $\gamma\in(0,1)$ in the ``pushed'' case. These corrections are the same  as those for the cut-off correction in travelling waves.

The cutoff result is known to give the actual correction in the pulled case to the velocity in the stochastic Fisher-KPP equation or in the $N$-BBM \cite{MuellerMytnikQuastel.2008,BerardGouere.2010}; it would be interesting to check whether the cut-off results of \autoref{sec:cut-off} also give the correction to the absorption rate~$\kappa$ in a stochastic $N$ particle Fleming-Viot process. It would also be interesting to explore the statistical properties of $Q_t/N$ for large but finite $N$. In particular, if the parallel to travelling waves still holds, one might expect that the variance of $Q_t/N$ be of order $(\ln N)^{-3}\,t$ in the pulled case. One could also try to understand the genealogies in Fleming-Viot and compare them to genealogies in the $N$-BBM~\cite{BrunetDerridaMuellerMunier.2006,BrunetDerridaMuellerMunier.2006a}.

 Another question would be to see if the large deviations of $Q_t/N$ are similar to the ones for the position  in the noisy Fisher-KPP or the $N$-BBM \cite{ElgartKamenev.2004,MeersonSasorov.2011,MeersonVilenkinSasorov.2013,MeersonSasorov.2024,DerridaShi.2016}. In that respect, let us mention that   we have been able to obtain an exact expression related to one particular value of the large deviation function of $Q_t/N$; let
\begin{equation}
V_j(t) = \frac 1N \Big\langle n_j(t) \Big(1-\frac 1N\Bigr)^{Q_t}\Bigr\rangle,\qquad S(t)=\Big\langle \Big(1-\frac 1N\Bigr)^{Q_t}\Bigr\rangle,
\end{equation}
where $n_j(t)$ is the number of particles at site $j$ and time $t$ in the Fleming-Viot system with $N$ particles. Assuming that, in the initial condition, the $N$ particles are independently distributed according to some $v_j(0)$, we can show (see \autoref{app:rel}) that the evolution of $V_j(t)$  satisfies the same equation \eqref{dtvj} as $v_j(t)$ with the same initial conditions. Therefore
\begin{equation}
\label{relconcl}
V_j(t)=v_j(t),\qquad S(t)=e^{-q_t},
\end{equation}
for any $t>0$ and any $N\ge2$.  As a consequence, if the initial condition decreases fast enough at infinity (for example, if the $N$ particles start from the same position $i_0$, that is $v_j(0)=\delta_{j,i_0}$), one obtains as $t\to\infty$
  for any $N\ge2$:
\begin{equation}
-\ln \Big\langle \Big(1-\frac 1N\Bigr)^{Q_t}\Bigr\rangle \sim \kappa_c t,
\label{123}
\end{equation}
where $\kappa_c$  is the critical absorbing rate given in~\eqref{eq:kappac}. It is remarkable that \eqref{123} does not depend on $N$ while the absorption rate  $\kappa_N =\lim_{t\to \infty} \langle Q_t\rangle/(N t)$  does.

\medskip

 We thank Pablo Ferrari and Baruch Meerson for useful discussions at an early stage of the present work.

\renewcommand{\theequation}{\thesection\arabic{equation}}
\def\theHequation{\thesection\arabic{equation}}
\setcounter{equation}{0}

\appendix
\section{Brownian motion with drift, absorption and reflection}
\label{app:Bdar}

Equation~\eqref{PP4} describes the evolution of the density of probability $v(x,t)$ that some Brownian motion on the positive half line with drift $-\mu$ with $\mu>0$ (so that the Brownian goes towards the origin) is at $x$ at time~$t$:
\begin{equation}
\partial_t v =\frac12\partial_ x^2 v +\mu \partial_x v\text{\quad for $x>0$},\qquad
\frac12 \partial_x v(0,t) = (a-\mu) v(0,t).
\label{PP4x}
\end{equation}
The meaning of the unusual boundary condition is that the Brownian motion is reflected at the origin, with some chance proportional to $a$ of being absorbed when at the origin. In this appendix, we try to explain this statement.

First, rather than having a Brownian motion confined on the positive half line and reflected at the origin, we prefer to think of a Brownian motion $B_t$ on the full half line with a drift $-\mu$ if $B_t>0$ and $+\mu$ if $B_t<0$. Then, calling $\tilde v(x,t)$ the density of probability for that Brownian motion, and assuming that $\tilde v(x,0)=v(x,0)$ is supported on the positive half line, one has for $x>0$:
\begin{equation}
v(x,t) = \tilde v(x,t) + \tilde v(-x,t).
\label{AA81}
\end{equation}

We now focus on $\tilde v(x,t)$. As a diffusion with absorption at the origin, it must follows the continuity equation of mass with a sink term proportional to $\delta(x)$.
\begin{equation}
\partial_t \tilde v + \partial_x j = - 2a\delta(x),
\label{AA82}
\end{equation}
where the coefficient $2a$ will turn out to be the correct coefficient to recover \eqref{PP4x}, and
where $j$ is the current of probability, related to $\tilde v$ by Fick's law with drift towards the origin:
\begin{equation}
j=-\frac12\partial_x \tilde v  - \mu \sign(x) \tilde v.
\label{AA83}
\end{equation}

Using \eqref{AA83} into \eqref{AA82}, and with $\partial_x\sign(x)=2\delta(x)$, we obtain that $\tilde v$ is solution to:
\begin{equation}
\partial_t \tilde v  = \frac12\partial_x^2 \tilde v +\mu\sign(x) \partial_ x\tilde v
+2(\mu-a)\delta(x)\tilde v.
\label{AA84}
\end{equation}
In this equation, $\tilde v$ is twice differentiable with respect to $x$ for any $x\ne0$. Moreover, $\tilde v$ is continuous at $x=0$ and,  making a formal integration  from $-\epsilon$ to $+\epsilon$ and taking $\epsilon\to0$, one has $0=\frac12\big[\partial_x \tilde v(0^+,t) - \partial_x \tilde v(0^-,t) \big] +2(\mu-a) \tilde v(0,t)$. Using \eqref{AA81}, this is the same boundary condition as in \eqref{PP4x}.

The solution to \eqref{AA84} can be obtained, as usual, by writing 
\begin{equation}
\tilde v(x,t)=\int_0^\infty\diffd y\,\tilde \pi(x,y,t) v(y,0),
\end{equation}
where $\tilde \pi(x,y,t)$ follows the same equation \eqref{AA84} as $\tilde v$ with the initial condition $\tilde \pi(x,y,t)=\delta(x-y)$. To find the explicit expression of $\tilde \pi(x,y,t)$, it is useful to make the following transformation:
\begin{equation}
\tilde \pi(x,y,t)=e^{-\frac{\mu^2}2t-\mu(|x|-|y|)}\psi(x,y,t) 
\label{AA85}
\end{equation}
Then
\begin{equation}\begin{aligned}
\partial_x\tilde \pi&= e^{-\frac{\mu^2}2t-\mu(|x|-|y|)}\big[-\mu\sign(x)\psi+\partial_x\psi\big]
,\\
\partial_x^2\tilde \pi&= e^{-\frac{\mu^2}2t-\mu(|x|-|y|)}\big[\big(\mu^2-2\mu\delta(x)\big)\psi-2\mu\sign(x)\partial_x\psi+\partial_x^2\psi\big]
\end{aligned}
\end{equation}
and we obtain that $\psi$ is solution to
\begin{equation}
\partial_t \psi = \frac12\partial_x^2\psi -b\delta(x)\psi,\qquad
\psi(x,y,0)=\delta(x-y),\qquad\text{with $b=2a-\mu$}.
\label{AA87}
\end{equation}
For the probabilist reader, the solution to \eqref{AA87} can be written as an expectation over Brownian motion using Feynman-Kac's formula:
\begin{equation}
\psi(x,y,t)\,\diffd x =\E_y\big[\indic{B_t\in\diffd x}e^{-b\int_0^t\diffd s\,\delta(B_s)}\big]=\E_y\big[\indic{B_t\in\diffd x}e^{-b\ell_t}\big],
\label{FK}
\end{equation}
 where $B_{s;s\ge0}$ is, under $\E_y$, a standard driftless Brownian started from $y$, and $\ell_t$ is the local time spent by $B$ at 0.
The expectation~\eqref{FK} is given in  \cite[{}1.1.3.7]{BorodinSalminen.2002}:
\begin{equation}
\psi(x,y,t)=
\frac1{\sqrt{2 \pi t}}{e^{-\frac{(x-y)^2}{2 t}}}
-bR\big(|x|+|y|,t\big),
\label{AA88}
\end{equation}
with
\begin{equation}
\begin{aligned}
R(X,t)&=
\frac12e^{\frac{b^2t}2+bX}\erfc\bigg(\frac{b\sqrt t}{\sqrt 2}+\frac X{\sqrt{2t}}\bigg)
=
\frac{1}{\sqrt{2\pi t}}\int_0^\infty\diffd z\,e^{-bz}e^{-\frac {(z+ X)^2}{2t}}.
\end{aligned}
\label{AAdefR}
\end{equation}

One can also check directly using the second form of \eqref{AAdefR} that \eqref{AA88} is indeed the solution to \eqref{AA87}.
Then, writing $\pi(x,y,t)=\tilde \pi(x,y,t) +\tilde \pi(-x,y,t)$, we recovers \eqref{PP6} using  \eqref{AA85}, \eqref{AA88}  and~\eqref{AAdefR}.

\setcounter{equation}{0}
\section{Derivation of the exact relation (\ref{relconcl})}
\label{app:rel}
\def\rate#1#2{r_{#1\to#2}}

In this appendix, we derive~\eqref{relconcl}. In fact, this relation is valid
for a Fleming-Viot process on any graph with arbitrary jumping rates.

We consider a particle jumping on an arbitrary graph:
we label the nodes of the graph by $i=0,1,2,\ldots$, and write $\rate i j $ for the jumping rate from site~$i$ to site $j$. Site~0 is absorbing, meaning that $\rate 0 j=0$ for all $j$. We set $\rate i i =0$ for all $i$.

Let $v_j(t)$ the probability that a single particle is at site $j$ at time $t$. One has
\begin{equation}
\partial_t v_j = \sum_{i} \big[ \rate i j v_i -\rate j i v_j\big].
\label{dtvj}
\end{equation}
As in~\eqref{C3}, we write $e^{-q_t}$ for the probability that the particle is not yet absorbed at site~0, and $u_j(t)$ the probability that it is at site $j>0$ given it is not absorbed yet:
\begin{equation}
e^{-q_t}=\sum_{j>0} v_j(t),\qquad u_j(t) = e^{q_t} v_j(t).
\end{equation}

We build a Fleming-Viot process on that graph. There are now $N$ particles, still jumping with the same rates. Each time a particle hits 0, it is instantaneously relocated at the position of one of the $N-1$ remaining particles. We write $n_j(t)$ for the number of particles at site~$j>0$ at time $t$, and $Q_t$ the number of absorptions (or jumps) up to time~$t$. We assume that, initially, the $N$ particles are independently distributed according to some initial density $u_i(0)=v_i(0)$. Then~\cite{FerrariMaric.2007}, in the large~$N$ limit, one has for any fixed~$t$:
\begin{equation}
\frac{n_j(t)}N\simeq u_j(t),\qquad \frac{Q_t}N\simeq q_t.
\end{equation}

For $N$ finite, one can write an evolution equation for $\langle n_j(t)\rangle$, but it involves two points correlations $\langle n_i(t) n_j(t)\rangle$ as in \eqref{eq:FV}. However, as we show now,
one can write a closed equation for $\langle n_j(t)A^{Q_t}\rangle$ for a well-chosen constant $A$.

\def\X#1#2{X_#1^{\scriptscriptstyle(#2)}}
We assign a label $a=1,2,\ldots,N$ to each particle and we let
$\X j a(t)$ be 1 if particle~$a$ is at position $j$ at time $t$ and 0 otherwise.
One has, for any $A$,
\begin{equation}
A^{Q_{t+\diffd t}} \X ja(t+\diffd t)=
\begin{cases}
A^{Q_t}&\text{proba. }\X j a\big(1-\sum_{i>0} \rate j i \,\diffd t-\sum_i n_i \rate i 0\, \diffd t\big)
\\&\qquad\qquad\qquad
+\sum_i \rate i j \X ia \, \diffd t,\\
A^{Q_t+1}&\text{proba. }\sum_i \rate i 0 \X ia\frac{n_j-\delta_i^j}{N-1} \, \diffd t,\\
&\qquad\qquad\qquad+\X j a \sum_i (n_i-\delta_i^j) \rate i 0 \,\diffd t\\
0 & \text{proba. everything else.}
\end{cases}
\end{equation}
(All the $n_i$ and $\X i a$ on the right hand side are taken at time $t$.
First case: to obtain $A^{Q_t}$, either particle $a$ is already on $j$, does not move and no other particle gets absorbed (first line), or particle $a$ jumps to $j$ (second line). Second case: to obtain $A^{Q_t+1}$, either particle $a$ jumps to 0 and is relocated on $j$, or it was already on $j$ and another particle jumps to~0.)

We replace the $\sum_{i>0}\rate j i$ term on the first line by  $\big(\sum_i\rate j i\big)-\rate j 0$. Two other $\rate j 0$ terms come from the $\delta_i^j$ in the sums. Then, we obtain,
\begin{equation}
\begin{aligned}
\partial_t\langle A^{Q_t}\X j a\rangle&=\biggl\langle A^{Q_t}\biggl[
\sum_i \bigl(\rate i j \X ia - \rate j i \X j a\bigr)
+ \rate j 0 \X j a \Bigl(1-\frac{AN}{N-1}\Bigr)
\\&\qquad\qquad
+\sum_i \rate i 0 \biggl(A\frac{\X i a n_j}{N-1}+(A-1)\X j a n_i\biggr)
\biggr]\biggr\rangle.
\end{aligned}
\end{equation}
We sum over $a$, using $n_j=\sum_a\X j a$,
\begin{equation}
\begin{aligned}
\partial_t\langle A^{Q_t}n_j\rangle&=\biggl\langle A^{Q_t}\biggl[
\sum_i \bigl(\rate i j n_i - \rate j i n_j\bigr)
+ \rate j 0 n_j \Bigl(1-\frac{AN}{N-1}\Bigr)
\\&\qquad\qquad
+\sum_i \rate i 0 \biggl(A\frac{n_i  n_j}{N-1}+(A-1)n_j  n_i\biggr)
\biggr]\biggr\rangle,
\\&=
\biggl\langle A^{Q_t}\biggl[
\sum_i \bigl(\rate i j n_i - \rate j i n_j\bigr)
+\Bigl(\frac{AN}{N-1}-1\Bigr)\sum_i \rate i 0 n_j  (n_i-\delta_i^j)
\biggr]\biggr\rangle.
\end{aligned}
\end{equation}
In particular, taking $A=1$ gives \eqref{eq:FV} for the model of \autoref{sec:lattice}.

Let us now choose $A=\frac{N-1}N$ and define
\begin{equation}
V_j(t) = \biggl\langle \frac{n_j(t)}N \Big(\frac{N-1}N\Big)^{Q_t}\biggr\rangle.
\end{equation}
We see that $V_j(t)$ follows the same equation \eqref{dtvj} as $v_j(t)$. Assuming that the initial position of the single walker has the same distribution as the initial position of each particle in the Fleming-Viot so that $V_j(0)=v_j(0)$, we obtain
\begin{equation}
V_j(t)=v_j(t),
\end{equation}
for any $t$, and any $N\ge2$. Taking the sum over $j$ gives
\begin{equation}
\biggl\langle \Big(\frac{N-1}N\Big)^{Q_t}\biggr\rangle= e^{-q_t}.
\label{143}
\end{equation}

In the particular case of the model of \autoref{sec:lattice}, the cutoff theory predicts that  $\langle Q_t\rangle/N$ grows for large $N$ with a rate $\kappa_c + \smash{\frac{C}{(\ln N)^2}}$ for any initial condition. However, \eqref{143} shows that $-\ln\langle e^{-a_NQ_t/N}\rangle$ (with $a_N=-N\ln(1-1/N)\simeq1$) grows, like~$q_t$, with a rate $\kappa$ which depends on the initial condition. This is because the expectation in \eqref{143} is dominated by rare events where there is a particle far away from the origin in the initial condition.

To conclude this appendix, we remark that other relations between higher order correlations exist. For instance, if we let
\begin{equation}
u_{jk}
= \biggl\langle\biggl(\frac{N-1}{N+1}\biggr)^{Q_t}\ \frac{n_jn_k-\delta_j^kn_j}{N(N-1)}\biggr\rangle,
\end{equation}
then one obtain after some algebra that $u_{jk}$ is solution to
\begin{equation}
\partial_t u_{jk} = \sum_i \Bigl(\rate i k  u_{ji} + \rate i j u_{ik} - (\rate j i+\rate k i) u_{jk}\Bigr)
+\frac{2\delta_j^k}{N+1} \sum_i \rate i 0  u_{ij}.
\end{equation}
This equation is closed and could be solved at least numerically. Then, one could obtain
\begin{equation}
\sum_{j,k} u_{j,k} = \biggl\langle\biggl(\frac{N-1}{N+1}\biggr)^{Q_t}\biggr\rangle.
\end{equation}

\bigskip

\paragraph{Conflict of interest} All authors have no conflicts of interest.
\paragraph{Data availability} No data was used for the research described in the article.

\printbibliography

\end{document}